\documentclass[aps,prd,superscriptaddress,showpacs,showkeys]{revtex4}
\def\be{\begin{equation}}
\def\ee{\end{equation}}
\def\bea{\begin{eqnarray}}
\def\eea{\end{eqnarray}}
\def\ba{\begin{array}}
\def\ea{\end{array}}
\def\e{{\rm e}}
\def\l{\lambda}

\begin{document}
\preprint{}
\title{
Correlation functions of the BC Calogero--Sutherland model}
\author{Shinsuke M. Nishigaki}
\email{shinsuke@phys.uconn.edu}
\affiliation{Department of Physics, University of Connecticut,
Storrs, CT 06269, USA}
\author{Dimitri M. Gangardt}
\email{gangardt@lkb.ens.fr}
\affiliation{LKB, Laboratoire de Physique de l'ENS, 24 rue Lhomond,
75005 Paris, France}
\author{Alex Kamenev}
\email{kamenev@physics.umn.edu}
\affiliation{Department of Physics, University of Minnesota,
Minneapolis, MN 55455, USA}
\date{September 4, 2002}
\begin{abstract}
The BC--type Calogero--Sutherland model (CSM) is an integrable
extension of the ordinary A--type CSM that possesses  a reflection
symmetry point. The BC--CSM is related to the
chiral classes of random matrix ensembles (RMEs) in exactly the
same way as the A--CSM is related to the Dyson classes. We first
develop the fermionic replica $\sigma$--model formalism suitable
to treat all chiral RMEs. By exploiting
``generalized color--flavor transformation''  we then extend the
method to find the exact asymptotics of the BC--CSM density
profile.
Consistency of our result with the  $c=1$
Gaussian conformal field theory description is verified.
The emerging Friedel oscillations structure and sum rules
are discussed in details.
We also compute the distribution of the particle nearest to the
reflection point.
\end{abstract}
\maketitle

\begin{center}
{\it To the memory of Prof.\ Sung-Kil Yang}\\
\ \\
\end{center}
\section{Introduction}
\label{s1}

It is a well known fact that there is an intimate relation between
the one--dimensional quantum problem with the inverse--square
interaction potential, i.e.\ Calogero--Sutherland model (CSM)
\cite{Cal,Sut}, and Dyson random matrix ensembles (RMEs)
\cite{Meh}. On the most elementary level the correspondence goes
as follows: for the three particular values of the coupling
constant ($\lambda = 1/2,1,2$) square of the many--body
ground--state wave function of the CSM coincides with the joint
probability distribution (JPD) of  RMEs  with Dyson index
$\beta=2\lambda$. Consequently the knowledge of the RME
correlation functions may be immediately translated to the
information about the CSM. Historically this correspondence proved to
be very fruitful for  advancing the understanding of the CSM.

It was later realized that Dyson's classification of the
RMEs was not exhaustive.
Studies of the two--sublattice model, mesoscopic transport, and
QCD Dirac spectra initiated the introduction of their ``chiral"
counterparts \cite{Gad,SN,SV,ForNPB,Ver}.
Subsequently, based on Cartan's
classification of Riemannian symmetric spaces,
Altland and Zirnbauer \cite{AZ,Zir2,AZ97} have further added six
``superconducting" chiral symmetry classes.
These chiral classes are characterized by
the special role played by the zero energy. Namely, there is a
certain number of eigenvalues that {\em must} have zero energy,
while all other eigenvalues  occur in symmetric pairs (mirror
images) around zero. As a result, the mean density of states (DoS)
exhibits either a hollow or a bump around the zero energy followed by
decaying oscillations at larger distances. Such a structure on the
level of the mean DoS does not show up in the Dyson classes and,
as we shall explain below, may be called Friedel oscillations.

The question is whether one can find an appropriate generalization
of the CSM whose  ground--state wave function possess the same
reflection symmetry property as the JPD  of chiral ensembles. The
answer is known to be affirmative. Indeed, one can write down an
integrable one--dimensional model with inverse--square
interaction, reflection symmetry and special single--particle
potential centered at zero having the required ground state. It is
known as the BC--type CSM in the literature \cite{OP}. Due to the
presence of the mirror boundary and localized single--particle
potential (an impurity) the model lacks translational invariance
and the resulting ground--state density is not uniform. In
particular the density profile develops the Friedel oscillations
far enough from the impurity. Accordingly one has a unique example
of integrable strongly interacting models that exhibit  Friedel
phenomena. That gives one a possibility to gain the exact
information on the amplitude decay rate, spectral characteristics
and phase shifts of the Friedel oscillations in the  interacting
system. Despite the proven integrability, explicit form of the
correlation functions of the BC--CSM with generic  coupling
constant were not established, except for some partial results
\cite{ForJMP,YKY2,AZ97,FM}.

The purpose of this paper it to fill this gap for rational values
of the coupling constant. To this end we employ the recently
developed approach based on the replica trick \cite{KM1,YL,NK}. It
was previously tested on the pair correlation function of the
ordinary A--CSM \cite{GK}, where it perfectly agrees with the
exact results of Haldane \cite{Hal} and Ha \cite{Ha} for any
rational coupling constant. The idea is to explore the relation
with the RMEs, where the replica trick was found to be accurate in
the asymptotic regime. We thus develop first the fermionic replica
approach to the chiral symmetry classes of RMEs. Not surprisingly,
we are able to reproduce the asymptotic behavior of the known DoS
profiles for chiral RMEs. We then extend the treatment away from
the RMEs values of the coupling constant and obtain  closed
analytic results for any rational two--body coupling constant
$\lambda$ and any impurity phase shift.

We found that for rational values of $\lambda=p/q$ ($p,q$
coprime), the spectrum of the Friedel oscillations contains
exactly $p$ harmonics, corresponding to $2k_F, 4k_F,\ldots ,
2pk_F$ density oscillations. The $l$--th harmonic ($l=1,\ldots,
p$) decays algebraically as $\theta^{-l^2/\lambda}$, where
$\theta$ is the distance from the impurity.  The amplitudes of the
harmonics depend on the number, $l$, and the coupling constant,
$\lambda$, but are not sensitive to the strength and details of
the impurity expressed through a phase shift, $\nu$. Moreover, the
harmonics amplitudes are closely related to those of the
two--point correlation function of the homogenous A--CSM. We
provide a conformal field theory account for this fact. The
impurity phase shift, $\nu$, affects only phases of the harmonics
in the asymptotic regime. We explicitly show that the relation
between the total charge expelled by the impurity and the
asymptotic phase shifts, known as the Friedel sum rule, holds for
the interacting system. We also compute the distribution of the
locus of the particle nearest to the mirror boundary point for the
BC--CSM at generic values of the coupling constants.

The paper is organized as follows: in section \ref{s2} we briefly
introduce chiral ensembles, and develop the appropriate fermionic
replica $\sigma$--models. Section \ref{s3} is devoted to the
introduction and replica treatment of the BC--CSM. In section
\ref{s4} we perform the analytical continuation and the replica
limit, and extract the  density profile. In section
\ref{sconf} we compare this result with the effective conformal field theory
description.
In section \ref{sEs} we compute the nearest particle distribution.
A summary and discussions are provided in section \ref{s5}.
Technicalities of the calculations are relegated to the two
appendices.

\section{Chiral Circular Ensembles}
\label{s2}

A  {\em circular} RME is defined as an ensemble of unitary
matrices $U$ representing a Riemannian symmetric space $\mathcal{D}$,
stochastically distributed according to the Haar measure $dU$ of
$\mathcal{D}$. Expressing a symmetric space as a coset ${\cal
D}={G}/{H}$ with a compact Lie group $G$ and $H\subset G$,
the Cartan mapping $G\to G/H$, $g\mapsto U(g)$
is as shown in the table below:
\be \ba{|l|l|l|l|} \hline
\mbox{class} &  G & H &U(g)\\
\hline
{\rm A\ (CUE)}& {\rm U}(N) & 1 & g \\
{\rm AI\ (COE)} & {\rm U}(N) & {\rm O}(N)& g^{T} g  \\
{\rm AII\ (CSE)} & {\rm U}(2N)&{\rm Sp}(2N)& g^{T} J g\\
{\rm AIII\ (chCUE)}&{\rm U}(N+N') &{\rm U}(N)\times{\rm U}(N')&
I g^{\dagger} I g  \\
{\rm BDI\ (chCOE)}& {\rm SO}(N+N')&{\rm SO}(N)\times{\rm SO}(N')&
I g^{T} I g\\
{\rm CII\ (chCSE)}
& {\rm Sp}(2(N+N'))& {\rm Sp}(2N)\times{\rm Sp}(2N')
& I g^{D} I g\\
{\rm D, B}& {\rm SO}(2N), {\rm SO}(2N+1) & 1 &
g\\
{\rm C}& {\rm Sp}(2N) & 1& g\\
{\rm CI}& {\rm Sp}(4N)& {\rm U}(2N)& I g^{\dagger} I g\\
{\rm DIII}_{e,o}
& {\rm SO}(4N),{\rm SO}(4N+2) & {\rm U}(2N),{\rm U}(2N+1) &
g^{D} g\\
\hline \ea \label{defU} \ee
Here
\be J=\left[\ba{cc}0&-
\openone_{N}\\ \openone_{N}&0 \ea \right],\ \ \
I=\left[\ba{cc}\openone_{N}& 0 \\ 0 &-\openone_{N'} \ea \right]
\ee
($N\to 2N, N'\to 2N'$ for CII and  $N, N'\to 2N$ for CI), and
$g^{D}=Jg^T J^{-1}$ denotes the quaternion dual of the matrix $g$.
Out of these twelve
classes,
 last non--classical nine
possess chirality \cite{Zir1}, i.e.\ non--zero eigenphases appear in complex
conjugate pairs. The JPD of these non--zero eigenphases
  \be
  U=V
{\rm diag}(\e^{i \theta_1},\ldots,\e^{i \theta_N}, \e^{-i
\theta_1},\ldots,\e^{-i \theta_N},1,\ldots,1) V^\dagger \, ,
  \ee
for   the circular ensemble is given by \cite{ForJE}
$(0\leq\theta\leq\pi)$
  \bea
P(\theta_1,\ldots,\theta_N)\, d\theta_1 \ldots d\theta_N &=&
\prod_{i=1}^N \left(d\theta_i \sin^{c_1} \frac{\theta_i}{2}
\cos^{c_2} \frac{\theta_i}{2} \right) |\Delta_N(\cos
\theta)|^\beta
\label{JE0}   \\
&=& \prod_{i=1}^N \left(dy_i \,
y_i^{(c_1-1)/2}(1-y_i)^{(c_2-1)/2} \right)
|\Delta_N(y)|^\beta\, , \ \ \ y_i\equiv\sin^2(\theta_i/2)\, ,
\label{JE}
  \eea
where $\Delta_N$ is the Vandermonde determinant of
the rank $N$ and the constants $c_1,\, c_2$ and $\beta$ for all
nine chiral ensembles are given in the table below.
\be
\ba{|l|cccc|l|l|}
\hline
\mbox{class}  & c_1 & c_2 & \beta & \nu& {\cal
D}\mbox{ (circular RME)} & \mathcal{M}\mbox{ (F-replica
NL$\sigma$M)}\\
\hline
\mbox{AIII}\ & 2(N'-N)+1 & 1 & 2 & N'-N &
{\rm U}(N+N')/\left({\rm U}(N)\times{\rm U}(N')\right)
& {\rm U}(n) \\
\mbox{BDI}\ & N'-N & 0 & 1 & N'-N &
{\rm SO}(N+N')/\left({\rm SO}(N)\times{\rm SO}(N') \right)&
{\rm U}(2n)/{\rm
Sp}(2n) \\
\mbox{CII}\ & 4(N'-N)+3 & 3 & 4 & N'-N &
{\rm Sp}(2(N+N'))/\left({\rm Sp}(2N)\times{\rm Sp}(2N') \right)&
{\rm U}(2n)/{\rm
O}(2n)\\
\mbox{D}\ & 0 & 0 & 2 & -1/2&
{\rm SO}(2N)& {\rm SO}(2n)/{\rm U}(n) \\
\mbox{B}\ & 2 & 0 & 2 & 1/2&
{\rm SO}(2N+1)& {\rm SO}(2n)/{\rm U}(n) \\
\mbox{C}\ & 2 & 2 & 2 & 1/2&
{\rm Sp}(2N)& {\rm Sp}(2n)/{\rm U}(n) \\
\mbox{CI}\ & 1 & 1 & 1 & 1&
{\rm Sp}(2N)/{\rm U}(N) & {\rm Sp}(2n) \\
\mbox{DIII}_{e}\ & 1 & 1 & 4 & -1/2&
{\rm SO}(4N)/{\rm U}(2N) & {\rm SO}(2n) \\
\mbox{DIII}_{o}\ & 5 & 1 & 4 & 1/2&
{\rm SO}(4N+2)/{\rm U}(2N+1) & {\rm SO}(2n)\\
\hline \ea \label{table}
  \ee
The JPD Eq.\ (\ref{JE}) defined over $y\in [0,1]$ may be
comprehensively called Jacobi ensemble. The constant $\nu$,
defined as
  \be
\nu=\frac{c_1+1}{\beta}-1\, ,
                                    \label{nu}
                                    \ee
may be called  ``topological charge'', when the tangent
element of a random matrix $U$ in the first three cases is interpreted
as the QCD Dirac operator in even dimensions \cite{SV,Ver}.

The real characteristic polynomial, or so called the fermionic
replicated partition function, for these circular ensembles is
defined as
  \be
Z_{n,N}(\theta)= \int\limits_{\mathcal{D}}
dU\,\det(\e^{i{\theta\over 2}}- \e^{-i{\theta\over 2}} U)^{n}
\equiv\int\limits_{G} dg\,\det(\e^{i{\theta\over 2}} - \e^{-
i{\theta\over 2}} U(g))^{n} \, .
                                      \label{charpol}
                                      \ee
After the color-flavor transformation \cite{Zir1} and
the thermodynamic limit where $N\to \infty, \theta\to 0$ with
their product (denoted by the same $\theta$
for the sake of simplicity) fixed finite,
it takes the form,
  \be
Z_{n}(\theta) \equiv
\lim_{N\to\infty}Z_{n,N}\left(\frac{\theta}{N}\right)=
{\theta}^{n\nu} \! \int\limits_{\mathcal{M}} \! du\,(\det u)^{\nu}\,
\e^{i\frac{{\theta}}{2}{{\,{\rm tr}\,}}(u+u^\dagger)}\, ,
                                                  \label{aftercft}
                                                  \ee
where $\mathcal{M}$ are the `dual' symmetric spaces of the fermionic
nonlinear
$\sigma$--models which are listed in the table (\ref{table}).
Hereafter we suppress irrelevant normalization constants, that
goes to unity in the replica limit, $n\to 0$. The derivation of
Eq.\ (\ref{aftercft}) from Eq.\ (\ref{charpol}) for the
AIII, BDI, and CII classes
is summarized in Appendix \ref{app1}. In order to derive
Eq.\ (\ref{aftercft}) for symmetry classes whose pertinent
color--flavor transformations are not immediately available in the
literature, one could as well
adopt an alternative method of magnifying the
origin of the circular ensembles first (i.e.\ employing Gaussian
ensembles), performing Hubbard-Stratonovich transformation and
taking the thermodynamic limit.
For a Gaussian ensemble treatment of the BDI and CII classes, see
Ref.\ \cite{HV}.

Performing the integration over angular degrees of freedom $v$ of
$u=v\,{\rm diag}(\e^{i\phi_a}) v^\dagger$, one obtains
  \be
Z_{n}(\theta) =\theta^{n\nu}\! \int\limits_0^{2\pi} \prod_{a=1}^n
\left(d\phi_a\,\e^{i\nu\phi_a+ i\theta\cos \phi_a}\right)
\left| \prod_{a>b}^n \sin\left(\frac{\phi_a-\phi_b}{2}\right)
\right|^{4/\beta}.
                                                     \label{nfold}
                                                     \ee
This expression is  valid for  all  nine chiral symmetry classes.
Note that Eq.\ (\ref{nfold}) depends only on $\beta$ and $c_{1}$
but not on $c_{2}$, because we have magnified the vicinity of the
origin, $\theta=0$. Accordingly the class C gives the same
$Z_{n}(\theta)$ as B, and the class CI the same as  BDI at
$c_{1}=1$, reducing nine symmetry classes of chiral RMEs to
seven universality classes. The number could be further  reduced
by introducing three universality classes of Laguerre ensembles,
having continuous $\nu$ and discrete $\beta=2,1,4$.

Consequently we succeeded in expressing the integration over the
initial $N$--variable  JPD, Eq.\ (\ref{JE0}), through the
$n$--fold integral, Eq.\ (\ref{nfold}). We shall discuss its
evaluation, analytical continuation and the replica limit after we
introduce the BC--CSM.
This replica treatment was previously performed
for the AIII class in Ref.\ \cite{DV}.

\section{BC Calogero-Sutherland Model}
\label{s3}

The generalized Calogero-Sutherland Hamiltonian \cite{OP}
is defined as
  \be
H=-\sum_{i=1}^N\frac{\partial^2}{\partial \theta_i^2}
+\sum_{\alpha\in \Delta}
\frac{g_{|\alpha|}}{\sin^2(\theta\cdot\alpha/2)}\, .
  \ee
Here $\Delta$ is a root system of a Lie algebra in
$N$--dimensional vector space, $\theta$ is the vector
$(\theta_{1},\ldots,\theta_{N})$, and $g_{|\alpha|}$ is a coupling
constant depending only on the root length. Quantum integrability
is ensured by these conditions. The ordinary, translationally
invariant model corresponds to the ${\rm A}_{N-1}$ root system.

The quantum one--dimensional model of $N$ interacting particles on
a semicircle, $0 \leq \theta_i \leq \pi$, where $i=1,\ldots, N$ with the
Hamiltonian
  \be
H  = -\sum_{i=1}^N\frac{\partial^2}{\partial \theta_i^2} \,+\,
\frac{\l(\l-1)}{2}\sum_{i>j}^N \left[
      \frac{1}{\sin^2 \frac{\theta_i-\theta_j}{2} }+
       \frac{1}{\sin^2 \frac{\theta_i+\theta_j}{2} }
\right]+\frac{\l_1(\l_1-1)}{4}\sum_{i=1}^N \frac{1}{\sin^2
\frac{\theta_i}{2}}\, +\, \frac{\l_2(\l_2-1)}{4}\sum_{i=1}^N
\frac{1}{\cos^2 \frac{\theta_i}{2}} \,
                                            \label{BCNCSM}
                                            \ee
corresponds to the ${\rm BC}_{N}$ root system, hence called the BC--CSM.
As the ${\rm BC}_{N}$ root system contains
roots of length $1,\sqrt{2}, 2$, there are three independent
coupling constants, $\lambda, \lambda_{1}, \lambda_{2}$.
This family contains CSMs corresponding to
${\rm B}_{N}$ ($\lambda_{2}=0$),
${\rm C}_{N}$ ($\lambda_{1}=\lambda_{2}$), and
${\rm D}_{N}$ ($\lambda_{1}=\lambda_{2}=0$)
root systems
as its subfamilies.
In addition to the pairwise inverse--square
interactions, the particles interact with their own mirror images
(occupying the other semicircle, $\pi \leq\theta \leq 2\pi$) and with
the two single--particle impurity potentials placed at $\theta =0$
and $\theta=\pi$. In what follows we shall assume the
thermodynamic limit, $N\to \infty$, and shall focus on the
vicinity of the impurity at $\theta = 0$. The other impurity may be
treated in exactly the same manner.
The model is known to have
the following ground--state energy \cite{OP}
  \be
H\Psi_0=E_0\Psi_0,\ \ \ \ \ \ \
\ E_0=\sum_{i=1}^N \left((N-i)\lambda +
\frac{\lambda_1+\lambda_2}{2}
\right)^2,
                                        \label{gse}
                                        \ee
and  the ground--state wave function
 \be
\Psi_0(\theta_1,\ldots,\theta_N)=
\prod_{i=1}^N \left( \sin^{\lambda_1} \frac{\theta_i}{2}
\cos^{\lambda_2} \frac{\theta_i}{2} \right)
\Delta_N(\cos \theta)^\lambda .
                                             \label{gswf}
 \ee
The absolute square of the ground--state wave function  coincides with
the JPD of the
chiral RMEs (\ref{JE0}) \cite{YKY},
through a change of the coupling constants
  \be
  \l=\beta/2\, , \,\,\,
\l_1=c_1/2\, ,\,\,\, \l_2=c_2/2\, .
                                          \label{constants}
 \ee
Notice, that  Eq.\ (\ref{gswf}) is {\em not} restricted to the
special values of $\beta, c_{1},  c_{2}$ listed in the
table~(\ref{table}).
The energy and wave functions of  the excited states
were studied in Refs.\ \cite{Yam,BPS,Ser,KO}.

The particle density is defined as,
                                      \be
\langle\rho(\theta)\rangle \equiv \left\langle \sum_{j=1}^N
\delta(\theta-\theta_j) \right\rangle=\sqrt{y(1-y)} \left\langle
\sum_{j=1}^N \delta(y-y_j) \right\rangle\, ,
                                       \label{density}
                                       \ee
where $y\equiv \sin^2(\theta/2)$. The angular brackets denote
ground--state expectation values, or equivalently averaging over
the normalized JPD, Eqs.\ (\ref{JE0}) and (\ref{JE}), for the first
and second equalities correspondingly. One may then employ the
replica trick to write
                                              \be
\sum_{j=1}^N\delta(y-y_j)= \lim_{n\to 0}\frac{1}{n\pi} \Im{m}
\frac{d}{dy} \prod_{j=1}^N (y-y_j-i\epsilon )^n\, .
                                         \label{replica}
                                         \ee
As a result, one obtains
                                      \be
\langle\rho(\theta)\rangle =\left. \sqrt{y(1-y)}\, \lim_{n\to 0}
\frac{1}{n\pi} \Im{m} \frac{d}{dy} Z_{n,N}(y-i\epsilon)
\right|_{y= \sin^2(\theta/2)}\, ,
                                          \label{replica1}
                                          \ee
where the ``replicated partition function'' is defined as
                                              \be
Z_{n,N}(y) = \int\limits_0^1 \prod_{i=1}^N \left( dy_i\,
y_i^{\l_1-1/2}(1-y_i)^{\l_2-1/2} (y-y_i)^n \right)
\left(\Delta_N(y)^2 \right)^{\lambda} \, .
                                                   \label{Zn}
                                                   \ee

Baker and Forrester \cite{BF} have noticed the integral  equality
due to Kaneko \cite{Kan} and Yan \cite{Yan},
which we suggestively call the ``generalized color-flavor
transformation''. With its  help  one may express the partition
function in the following way
                                     \be
Z_{n,N}(y) = \int\limits_{\mathcal{C}} \prod_{a=1}^n
\left(dx_a\,x_a^{ \frac{\l_1+\l_2+1}{\l}-2 }(1-x_a)^{
-\frac{\l_2+n-1/2}{\l} } \left[ \frac{x_a(1-y x_a)}{1-x_a}
\right]^N \right)
\left( \Delta_n(x)^2 \right)^{1/\l} \, .
                                       \label{gcftII}
                                       \ee
The integration contour $\cal{C}$ encircles the cut between
$x_a=0$ and $x_a=1$.
The general form of the integral identity is given in  Appendix
\ref{app2}.
So far no approximation has been made.
Now we pass  to the thermodynamic limit, $N\to \infty$, and
magnify the vicinity of the $\theta=0$ impurity. To this end we
rescale the variable as $\theta\to \theta/N$ and correspondingly
$y\simeq\theta^2/(4N^2)$. By redefining the integration variables
as $x_a=1-2iN\theta^{-1}\e^{i \phi_a}$ and taking the
thermodynamic limit, $N\to \infty$, one finds (see Appendix
\ref{app2} for details)
  \be
Z_{n}(\theta) \equiv \lim_{N\to\infty}
Z_{n,N}\left(\frac{\theta}{N}\right) =
\theta^{n(\frac{\l_1}{\lambda} +\frac{1}{2\lambda}-1)}
\int\limits_{0}^{2\pi} \prod_{a=1}^n \left(d\phi_a\,
\e^{i(\frac{\l_1}{\lambda} +\frac{1}{2\lambda}-1)
\phi_a+i\theta\cos\phi_a} \right) \left[\prod_{a>b}^n
\sin^2\left( \frac{\phi_a-\phi_b}{2}\right) \right]^{{1}/{\l}} .
                                              \label{nfold1}
                                              \ee
One may  introduce  notation
                                              \be
\nu = \frac{\l_1}{\lambda} + \frac{1}{2\lambda}-1\, ,
                                                \label{nu2}
                                                \ee
to notice the exact coincidence with the $\sigma$--model
representation of the chiral RMEs, Eq.\ (\ref{nfold}), provided
that the coupling constants are related via Eq.\
(\ref{constants}).
The important difference is that the BC--CSM representation in the
form of  Eq.\ (\ref{nfold1}) is {\em not} restricted to the RMEs
values $\lambda=1/2, 1, 2$ and special values of the topological
charge, $\nu$.

\section{Analytic Continuation and Replica Limit}
\label{s4}

Consider the $n$--fold integral
                              \be
Z_{n}(\theta) = \theta^{n\nu} \int\limits_0^{2\pi} \prod_{a=1}^n
\left(d\phi_a\,\e^{i\nu\phi_a+ i\theta \cos \phi_a}\right)
\left[ \prod_{a>b}^n
\sin^2\left( \frac{\phi_a-\phi_b}{2}\right)
\right]^{1/\lambda}\, .
                                             \label{nfold2}
  \ee
One may  stretch the integration contour  from  the unit circle
in the complex plane of $z_a=\e^{i\phi_a}$  into two lines
parallel to the imaginary axis with $\Re{e}\, z_a =\pm 1$. The
original integral, Eq.\ (\ref{nfold2}), splits into the sum of $n$
terms with $l$ integrals having $\Re{e}\, z_a = -1$ and remaining
$n-l$ ones $\Re{e}\, z_a = 1$; here $l=1,\ldots, n$. The further
progress is made possible in the asymptotic limit, $\theta\gg 1$.
In this case the integrals are dominated by the vicinities of the
saddle points $z_a=\pm 1$ and thus may be evaluated employing the
Selberg integral. This strategy was described in  details in
Refs.~\cite{KM1,YL,NK,GK}. Proceeding this way, one finds for the
replicated partition function
  \be
  Z_{n}(\theta)
=  \theta^{n\nu} \e^{in\theta} \sum\limits_{l=0}^n
F^l_n(\lambda)\, \e^{i\pi\nu l-2il\theta}\, 2^{-{2l^2\over
\lambda}}\, (i\theta)^{-{n-l\over 2}\left(1+{n-l-1\over
\lambda}\right)}\, (-i\theta)^{-{l\over 2}\left(1+{l-1\over
\lambda}\right)}\,\, ,
                                           \label{saddles}
  \ee
where we have omitted a normalization constant that goes to unity
in the replica limit, $n\to 0$ and, following Refs.\
\cite{KM1,YL,NK,GK}, introduced the notation
  \be
F^l_n(\lambda)\equiv
{n\choose l} \prod\limits_{a=1}^l\frac{ \Gamma(1+a/\lambda)
}{\Gamma(1+(n-a+1)/\lambda)} \, .
                                                \label{F}
  \ee
Employing the observation that $F^l_n(\lambda)\equiv 0$ for $l>n$,
one may  extend  summation over $l$ in Eq.\ (\ref{saddles}) to
infinity and then perform the analytic continuation $n\to 0$. As a
result, one finds via $ \langle\rho(\theta)\rangle= \lim_{n\to 0}
(\pi n)^{-1} \Im{m}\,
\partial_\theta Z_{n}(\theta)
$
(cf.\ Eq.\ (\ref{replica1})),
  \be
\langle\rho(\theta)\rangle = {1\over\pi} \left[1 + 2
\sum\limits_{l=1}^{\infty}
\frac{d_l(\lambda)}{(2\theta)^{l^2/\lambda}}\, \cos\left(
2l\theta-l{\pi} \left(\nu + \frac12-\frac{1}{2\lambda}\right)
\right)\right] \, ,
                                          \label{rhoRSB}
  \ee
where
  \be
  d_l(\lambda) \equiv \frac{(-1)^l}{2^{l^2/\lambda}}
  \prod\limits_{a=1}^l {
\Gamma(1+a/\lambda) \over \Gamma(1-(a-1)/\lambda)}\, .
                                             \label{dl}
  \ee
Eq.\ (\ref{rhoRSB}) for the asymptotic of the ground--state density
of the BC--CSM is the central result of this paper. For the
integer values of the coupling constant, $\lambda$, it may be
extracted from the expressions derived by Baker and Forrester
\cite{BF}, see also Ref.~\cite{FM}.

We remark that each term in Eq.\ (\ref{rhoRSB}) is the one
with the lowest power in $\theta^{-1}$ among all terms carrying
the same frequency $2l$. This corresponds to truncating
all contributions from the descendent fields
in the conformal description, see the next section.
One could compute these secondary terms, subleading in $\theta$,
by performing a perturbative expansion
around each saddle point with the help of the loop equations \cite{NK}.

\section{Conformal Field Theory Description}
\label{sconf}

By comparing the one--particle correlation function of the BC--CSM
(\ref{rhoRSB}) with the equal--time {\em two--particle}
correlation function of the ordinary A--CSM \cite{Ha,GK}
  \be
  \left\langle \rho(\theta) \rho(0) \right\rangle =
{1\over (2\pi)^2}\left[
1- {1 \over 2  \lambda \theta^2} +
2\sum\limits_{l=1}^\infty {d_l(\lambda)^2 \over
\theta^{2l^2/\lambda} } \, \cos(2 l \theta)\right] \, , \label{R1}
  \ee
one notices that it is expressed through the very same
coefficients $d_l(\lambda)$, Eq.\ (\ref{dl}). It is thus clear
that the harmonics amplitudes, $d_l(\lambda)$, are properties of
the homogenous interacting system and {\em not} of the localized
impurity. As explained below, this result could be anticipated
from the effective conformal field theory description. The latter
is capable of predicting the low energy properties of the system
apart from numerical values of the coefficients.

Based on the finite--size scaling analysis, Kawakami and Yang
\cite{KY} identified the low--energy effective theory of the CSM
in the thermodynamic limit to be the $c=1$ Gaussian conformal
field theory at radius $R=\sqrt{\lambda}/2$, either non--chiral
(A--CSM) \be
  \mathcal{L}=\frac{1}{2\pi}\partial_{z}\Phi\partial_{\bar{z}}\Phi\, ,
  \ \ \
\Phi(z,\bar{z})=\phi(z)+{\bar{\phi}}(\bar{z})\, , \ \ \
\Phi\equiv\Phi+2\pi{R} \, ,
                    \label{compact}
  \ee
or chiral (BC--CSM) \cite{YKY,YKY2}.
Namely, they have found the density operator
should have zero winding number.
It does not have a definite conformal dimension, and
therefore is expanded in terms of primary and secondary operators
whose left-- and right--moving vertex momenta are equal.
Here we shall consider only contributions from the ${\rm U}(1)$
current and the primary fields (vertex operators with charges
allowed by the compactification (\ref{compact})),
  \be
\rho(z,\bar{z})=\rho_0 \left[ b \left(
\partial_z \phi(z) +
\partial_{\bar{z}} {\bar{\phi}}
(\bar{z}) \right) +\sum_{l=-\infty}^\infty d_l \,
\e^{il(z+\bar{z})}
\e^{il\phi(z)/{R}}\e^{il{\bar{\phi}}(\bar{z})/{R}}
\right] \, ,
\label{rhoprimary}
  \ee
where the expansion coefficients, $b$ and $d_l(=d_{-l})$, are not
determined from the conformal field theory.
Neither are the oscillation factors $\e^{il(z+\bar{z})}$, which are
set by hand to describe
transport of $l$ pseudo--particles \cite{Ha} from the left Fermi point
($-k_F=-1$ by normalization)
to the right Fermi point ($k_F=1$) \cite{KY}.
We have factored out
$\rho_0$ such that the constant term carries $d_0=1$. The
propagator and the vertex correlator are given by
  \be
  \left\langle \phi(z)\phi(z') \right\rangle =
-\frac14 \log(z-z')\, ; \hskip 1cm
\left\langle\e^{il\phi(z)/{R}}\e^{il'\phi(z')/{R}}\right\rangle
  =
\frac{\delta_{l,-l'}}{(z-z')^{l^2/(4{R}^2)}}\, .
                             \label{rhorhoA}
  \ee
The coefficient in the second equation is a matter of convention and
reflects a particular choice of the ultraviolet regularization.
Another choice of the regularization would change coefficients
$d_l$, but not the final result. Employing
Eqs.\ (\ref{rhoprimary}), (\ref{rhorhoA}),  one obtains for the
equal--time two--particle correlation function of the A--CSM (we
denote $z=\theta+i \tau$)
  \be
  \langle \rho(\theta+i
0)\rho(\theta'+i 0) \rangle =\rho_0^2 \left[
-\frac{b^2}{2(\theta-\theta')^2} + \sum_{l=-\infty}^\infty d_l^2\,
\frac{\e^{2il(\theta-\theta')}}{(\theta-\theta')^{l^2/(2{R}^2)}}
\right].
  \ee
Comparing this expression with Eq.\ (\ref{R1}), one finds that the
$b=1/\sqrt{\lambda}$ \cite{Ha}, while $d_l=d_l(\lambda)$, cf.
Eq.\ (\ref{dl}). Let us consider now the BC--CSM  and concentrate
on the case without the phase shift for simplicity.
The Dirichlet
boundary condition at $\Re{e}\, z=0$
 is translated into the open boundary bosonization rule \cite{GNT}
\be
\phi(z)=-{\bar{\phi}}(z) \ \ \ \mbox{for}\ \ \ \Im{}m\,z<0.
\label{LR}
  \ee
As the right mover is identified as $(-1)$ times the left mover at
the mirror--imaged point, there exists a nonzero matrix element
between the left-- and right--moving vertex operators:
 \be
\left\langle \e^{il\phi(z)/{R}}\e^{il'{\bar{\phi}}
(\bar{z}')/{R}}\right\rangle =
\frac{\delta_{l,l'}}{(z+\bar{z}')^{l^2/(4{R}^2)}}\, .
  \ee
As a result, the mean density of the BC--CSM becomes
nontrivial,
  \be
\langle \rho(\theta+i 0)\rangle =\rho_0 \sum_{l=-\infty}^\infty
d_l \frac{\e^{2il\theta}}{(2\theta)^{l^2/(4{R}^2)}}\, .
\label{rhoBC}
  \ee
It is in  exact agreement with Eq.\ (\ref{rhoRSB}) if one
disregards the phase shift. The phase shift may also be included in the
conformal description by shifting the identification of the right
and left movers, Eq.\ (\ref{LR}), by a constant factor: $- \pi\left
(\nu+1/2-1/(2\lambda)\right)$.

We found that the knowledge of the asymptotic behavior of the
correlator of the homogenous A--CSM, Eq.\ (\ref{R1}), supplemented
by the conformal field theory description is, in principle,
sufficient to predict the BC--CSM correlation function,
Eq.\ (\ref{rhoRSB}). This agreement indicates the fact that the
ultraviolet property of the field that is responsible for the
normalization  of vertex operators are not  affected by the
presence or absence of the boundary. Well anticipated as it is, we
nevertheless consider this fact to be worth verifying, as done in
this paper. This fact can be put on a further test by computing
the asymptotics of e.g.\ two--particle correlation function for
the BC--CSM, though it is technically more challenging.
We could as well reverse the logic and conjecture the asymptotically
expanded form of
any $p$-point correlation function of density operators
for the A--CSM ($p\geq 3$) or
for the BC--CSM ($p\geq 2$), by using
Eqs.\ (\ref{rhoprimary}), (\ref{rhorhoA}), and (\ref{dl}).

\section{Nearest Particle Distribution}
\label{sEs}

Another quantity of interest in the theory of interacting
electrons is the probability $E[s',s]$ of finding no particle
within an interval $[s',s]$ or the particle spacing distribution
$p(s)$ that is a derivative of the former. In the context of spin
chains a similar quantity was recently discussed in
Ref.~\cite{Abanov}. For nonchiral as well as chiral RMEs, Tracy
and Widom \cite{TW} has developed  Mehta's computation \cite{Meh}
of $E[s',s]$ as a Fredholm determinant into a systematic and
powerful method.  As their method  determines $E[s',s]$ as a
solution ($\tau$ function) to a transcendental equation of
Painlev\'e type relies upon the orthogonal polynomials, its validity
is necessarily limited to $\lambda=1/2,1,2$.  On the other hand,
$E(s)\equiv E[0,s]$ for  the chiral RMEs has been computed by an
alternative and far simpler ``shifting'' method
\cite{SN,ForNPB,ForJMP,FH,DN}
as explained below.  We  show that with a help of the generalized
color--flavor transformation, this method is applicable also to
the BC--CSM at generic values of
the coupling constants.
Consider for a moment $\lambda_1=n+1/2$, where  $n=0,1,2,\ldots$
is an integer and $\lambda_2=1/2$. As we are interested in the
universal behavior in the vicinity of the reflection point
$\theta=0$, the restriction on $\lambda_2$ is irrelevant. The
probability of having no particle within an interval
$0\leq\theta\leq s$, or $0\leq y\leq Y$ with $Y=\sin^2(s/2)$, is
defined as
 \be
E_N(s)=\mathrm{const.} \int\limits_Y^1 \prod_{i=1}^N \left(
dy_i\,y_i^n \right) \left( \Delta_N(y)^2 \right)^{\lambda} \,.
 \ee
The constant should be chosen to ensure $E_N(0)=1$. By shifting
and rescaling the integration variable as $y\to (1-Y)y+Y$, one
obtains
 \be
 E_N(s)=\mathrm{const.}(1-Y)^{(1+n)N+\lambda N (N-1)}
\int\limits_0^1 \prod_{i=1}^N \left[
dy_i\left(y_i+\frac{Y}{1-Y}\right)^n \right] \left( \Delta_N(y)^2
\right)^{\lambda}\, .
 \ee
Now we apply the generalized color--flavor transformation
(\ref{gcftII}) to reexpress $E_N(s)$ as an $n$--fold integral,
 \be
E_N(s)=\mathrm{const.}(1-Y)^{(1+n)N+\lambda N (N-1)}
\int\limits_{\mathcal{C}} \prod_{a=1}^n \left[ dx_a\,
x_a^{\frac2\lambda -2} (1-x_a)^{-\frac{n}{\lambda}} \left[
\frac{x_a(1+\frac{Y}{1-Y}x_a)}{1-x_a} \right]^N \right]
(\Delta_n(x)^2)^{{1}/{\lambda} } \, .
 \ee
We finally rescale $s\to s/N\,; \,\,  Y\to s^2/(4N^2)$ and take
the thermodynamic limit. Following the same procedure that lead
from Eq.\ (\ref{gcftII}) to Eq.\ (\ref{nfold1}),  one finds
 \bea
E(s)&\equiv& \lim_{N\to\infty} E_N\left(\frac{s}{N}\right)
\nonumber\\
&=& \mathrm{const.}\,\e^{-\frac{\lambda}{4}s^2}
s^{n(1-\frac{1}{\lambda})} \int\limits_{0}^{2\pi} \prod_{a=1}^n
\left(d\phi_a\, \e^{i(\frac{1}{\lambda} -1) \phi_a+s\cos\phi_a}
\right)\left[ \prod_{a>b}^n \sin^2\left(\frac{\phi_a-\phi_b}{2}\right)
\right]^{{1}/{\l}} \,.
                                              \label{Es}
                                              \eea
This exact result has previously been derived from
Laguerre (chiral Gaussian) ensembles
at arbirtary $\lambda$ and at $\lambda_1=n+1/2$ \cite{ForJMP}.

To compute the asymptotics of $E(s)$ for $s\gg 1$,
one can relax the restriction on $\lambda_1$, by
first evaluating the $n$--fold
integral (\ref{Es}) by the saddle point method and
then performing the analytical continuation
$n\to \lambda_1 -1/2$.
In the large--$s$ limit we pick only the
contribution of the replica-symmetric saddle point $\phi_a = 0$,
since the contribution of all other saddle points is exponentially
smaller (the difference with the previous computation is that $s$
enters Eq.\ (\ref{Es}) without imaginary unit).
This way Forrester \cite{ForJMP} has derived
the asymptotic $(s\gg 1)$ result
 \bea
E(s) &= & \mathrm{const.\ }s^{\frac{n}{2}-\frac{n(n+1)}{2\lambda}}
\e^{-\frac{\lambda}{4}s^2}
\e^{ns}  \nonumber\\
&= &\mathrm{const.\ } s^{-\nu \left(\lambda_1-\frac{1}{2}\right)}
\e^{-\frac{\lambda}{4}s^2+\left(\lambda_1-\frac{1}{2}\right)s}\, ,
                                          \label{realsaddle}
 \eea
where $\nu$ is the topological charge defined by Eq.\ (\ref{nu2}).
The Gaussian factor $\exp( -\lambda s^2/4)$ could
be anticipated from the mean--field treatment of the classical
logarithmic gas \cite{Abanov}.
The other factors in the asymptotic expression
Eq.\ (\ref{realsaddle}) could not be found in any simpler way, to
the best of our knowledge.

The distribution $p(s)$ of the locus $s$ of the particle nearest
to the reflection point is given by $p(s)=-\partial_s E(s)$. In
the other limiting case, $s\ll 1$, $p(s)$ is determined by the
interaction of the particle closest to the reflection point to its
own mirror image. Inspecting Eq.\ (\ref{JE}) one immediately finds
$p(s) \propto s^{2\lambda_1}$.

\section{Discussions}
\label{s5}

Let us now take a closer look at our main result, Eq.\
(\ref{rhoRSB}). The constant term on its r.h.s., $\rho_0=1/\pi$,
represents the uniform density of particles ($N$ particles within
$[0,\pi]$) far away from the impurity. It may be traced back to
the replica symmetric contribution (all $n$ integrals are taken at
$z_a=-1$ saddle point) to the partition function. The decaying (as
$\theta\to \infty$) oscillatory terms on the r.h.s.\ of Eq.\
(\ref{rhoRSB}) are the Friedel oscillations of the particle
density induced by the mirror boundary and the impurity potential.
These terms may be identified as the replica symmetry broken
contributions to the partition functions ($l$ integrals are taken
at the ``wrong'' saddle point, $z_a=1$). In general, there is an
infinite number of harmonics (unlike a single ``$2k_F$'' harmonic
in the non--interacting system!)  in the oscillation spectrum.
Since $d_l\propto \exp( \lambda^{-1} l^2\ln l)$ for $l\gg 1$,
the sum over harmonics on the r.h.s.\ of Eq.\ (\ref{rhoRSB}) is, in
general, not convergent. It is not clear to us at the moment,
whether there is a consistent resummation scheme. There is,
however, an important class of the parameters, where
Eq.\ (\ref{rhoRSB}) is mathematically rigorous.

For any rational coupling constant $\lambda=p/q$ the coefficient
$d_l(p/q)\equiv 0$ for $l>p$ and therefore the sum terminates
after exactly $p$ oscillatory components. One finds that, in
addition to the usual $2k_F$ Friedel oscillation, the system
possess $4k_F,\ldots, 2pk_F$ oscillatory components of the density
(in the unit $k_F= 1$ accepted here). This fact might be
expected from the form of the density--density response function
of the homogenous A--CSM \cite{Hal,Ha}. However, the algebraic
decay rate of the harmonics could {\em not} be determined
employing linear response of the A--CSM. Indeed, the latter
predicts that the $l$--th harmonic decays as
$\theta^{-2l^2/\lambda+1}$, while the correct decay rate is
$\theta^{-l^2/\lambda}$. Notice, that for non--interacting
particles, $\lambda=1$ and thus $l=1$, both ways give the correct
one--dimensional decay of the $2k_F$ Friedel oscillations:
$\theta^{-1}$. For any interacting system, $\lambda\neq 1$, the
linear response is bound to fail in the asymptotic regime. These
observations was already made in the
Tomonaga--Luttinger liquid literature
\cite{KF,EG}. Now we can confirm them having the exactly solvable
model system.

In the BC--CSM one finds not only the decay law, but also the
relative amplitudes of the harmonics: coefficients $d_l(\l)$, Eq.\
(\ref{dl}). Notice that these amplitudes are determined by the
interaction strength, $\lambda$, only and are independent on the
impurity strength, $\lambda_{1}$. This is due to the fact that the
mirror boundary condition  induces oscillations of the maximal
possible amplitude. The additional single--particle potential
centered at $\theta=0$ and characterized by $\lambda_1$ changes
the phase of the oscillations only.
The entire
information about the impurity strength, $\lambda_{1}$, is
incorporated in the parameter $\nu$, Eq.\ (\ref{nu2}). In the
asymptotic regime the latter affects the phase of the Friedel
oscillations only and therefore may be associated with the
impurity phase shift. (Unlike the leading order, the amplitudes of
sub--leading perturbative corrections in negative powers of $\theta$
do depend on the phase shift,
$\nu$.)

To verify Eqs.\ (\ref{rhoRSB}), (\ref{dl}) one may compare them
with the available exact DoS of the chiral RMEs (see Ref.\
\cite{Iva} and references therein). Employing Hankel's
asymptotic expansion of the Bessel function, one may check that
these asymptotic  perfectly agree with Eqs.\ (\ref{rhoRSB}),
(\ref{dl}). One may also notice that for D, B, and C symmetry
classes we have obtained the exact rather than the asymptotic
results. This coincidence is  due to the Duistermaat--Heckman
localization theorem \cite{DH,Zir99}. We see that having unitary
symmetry class, $\beta=2$, is not sufficient to satisfy the
Duistermaat--Heckman theorem. One should also have the special
value of the topological charge, $\nu=\pm 1/2$, to secure
cancellation of all higher--order perturbative corrections
\cite{DV}.

One may define the total charge attracted (expelled) by the
impurity to (from) the region near $\theta=0$ as:
  \be
Q\equiv \int\limits_{0}^\infty d\theta\, (\rho(\theta) - \rho_0)
\, ,
                                               \label{Q}
  \ee
where $\rho_0=1/\pi$ is the uniform asymptotic density. For the
chiral RMEs, where the exact expressions including small
$\theta$ region are available, the result is (cf.
Eq.\ (\ref{rhoRSB}) ):
  \be
Q = - \frac{1}{2} \left(\nu+\frac12-\frac{1}{2\lambda}\right) =
\frac{1}{4} -\frac{\lambda_1}{2\lambda}  \, \, .
                                            \label{Friedelsum}
  \ee
Notice, that the attracted charge depends both on the impurity
amplitude, $\lambda_1$ and the interaction strength, $\lambda$,
reflecting the fact that the impurity is screened due to the
interactions.  The pure mirror boundary, without the
single--particle potential attracts  quarter of a particle
irrespective to the interaction strength.
Equation~(\ref{Friedelsum})  is a manifestation of the famous
Friedel sum rule: the total expelled charge is equal to the
impurity phase shift (divided by $2\pi$); the latter also
determines the phase of the density oscillations far from the
impurity. We conjecture, in accordance with the earlier works
\cite{Langer}, that Eq.\ (\ref{Friedelsum}) is valid for any values
of $\l$ and $\lambda_{1}$.

\begin{acknowledgments}
We are grateful to the hospitality of Department of Physics,
Technion--Israel Institute of Technology, Haifa, Israel,
where our collaborated work has been initiated.
We also acknowledge
A. Abanov, A. Altland, G. Dunne, A. Garc\'{i}a-Garc\'{i}a,
P. Forrester, J. Verbaarschot, and M. Zirnbauer
for valuable discussions and correspondences. 
This work was supported in part (SMN) by the
DOE grant no.\ DE-FR02-92ER40716, and (AK) by the BSF grant no.\ 
9800338. Le LKB est UMR 8552 du CNRS, de l'ENS et de Universit\'e
P.\ et M.\ Curie.
\end{acknowledgments}

\appendix
\section{$\sigma$--Model Derivation via Color--Flavor Transformation}
\label{app1}

We first consider the simplest case AIII with $\nu=0$.
The Grassmannian
${\rm U}(2N)/\left({\rm U}(N)\times{\rm U}(N)\right)$
is a complex
K\"{a}hler manifold, and its unitary matrix representative $U$
in Table (\ref{defU})
is conveniently parameterized by the complex
stereographic coordinate $Z_{ij}$, $i,j=1,\ldots,N$,
as
\be
U=I g^\dagger I g=I \gamma I \gamma^{-1},
\ \ \
\gamma=\left[\ba{cc} 1& Z\\
       -Z^\dagger  &1
\ea \right] ,
\ \ \ Z\in \mathbf{C}^{N\times N},
\ \ \
I=\left[\ba{cc} 1& 0\\
       0  & - 1 \ea \right].
\label{param}
\ee
The K\"{a}hler
potential $k(Z,\bar{Z})={{\rm tr}\,}\log(1+ZZ^\dagger)$ leads to the
Haar measure
\be
dU=\frac{\prod_{i,j=1}^N d^2
Z_{ij}}{\det(1+ZZ^\dagger)^{2N}}\, .
\label{meas}
\ee
The replicated partition function then reads
 \bea
Z_{n,N}(\theta)&=&
\int_{{\rm U}(2N)/\left({\rm U}(N)\times{\rm U}(N)\right)}
dU\,\det (\e^{i{\theta\over 2}} - \e^{-i{\theta\over 2}} U)^n
 \nonumber\\
&=&\int_{\mathbf{C}^{N\times N}} \frac{\prod d^2
Z}{\det(1+ZZ^\dagger)^{2N}} \det \left( \e^{i{\theta\over 2}} -
\e^{- i{\theta\over 2}} I \gamma I \gamma^{-1}
\right)^{n} \nonumber\\
&=& \int_{\mathbf{C}^{N\times N}} \frac{\prod d^2
Z}{\det(1+ZZ^\dagger)^{2N+n}} \det \left( \e^{i{\theta\over 2}}
\gamma - \e^{- i{\theta\over 2}} I \gamma I \right)^{n}\, .
 \eea
We introduce a set of $(N\times n)$--component independent
Grassmannian numbers $\psi^{a}_{i}, \chi^{a}_{i},
\bar{\psi}^{a}_{i}, \bar{\chi}^{a}_{i}, i=1,\ldots,N,
a=1,\ldots,n$ to exponentiate the determinant, ($I\gamma
I=\gamma^\dagger$)
 \be
Z_{n,N}(\theta)= \int {d}\psi d\bar{\psi}{d}\chi d\bar{\chi}
\int_{\mathbf{C}^{N\times N}} \frac{\prod d^2
Z}{\det(1+ZZ^\dagger)^{2N+n}} \, \exp\left( [\bar{\psi}\
\bar{\chi}] \left( \e^{i{\theta\over 2}} \gamma -
\e^{-i{\theta\over 2}} \gamma^{\dagger}\right) \left[ {\psi \atop
\chi} \right] \right)\, .
 \ee
Now we employ Zirnbauer's color--flavor transformation \cite{Zir1}
 \be
\int_{\mathbf{C}^{N\times N}} \frac{\prod d^2
Z}{\det(1+ZZ^\dagger)^{2N+n}} \exp\left( \bar{\psi}_i^a Z_{ij}
\chi_j^a - \bar{\chi}_i^a Z_{ij}^{\dagger} \psi_j^a \right) =
\int_{{\rm U}(n)} du\, \exp\left( \bar{\psi}_i^a u^{ab} \psi_i^b +
\bar{\chi}_i^a u^{\dagger}{}^{ab} \chi_i^b \right)\, ,
 \ee
to obtain
 \bea
Z_{n,N}(\theta)&=& \int {d}\psi d\bar{\psi}{d}\chi d\bar{\chi}
\int_{{\rm U}(n)} du\, \exp\left( (\e^{i{\theta\over 2}} -
\e^{-i{\theta\over 2}}) (\bar{\psi}\psi+\bar{\chi}\chi) +
(\e^{i{\theta\over 2}} + \e^{-i{\theta\over 2}})
(\bar{\psi}u\psi+\bar{\chi}u^\dagger\chi) \right)
\nonumber\\
&=&
\int_{{\rm U}(n)} du\,
\det\left(
\cos\theta +i\sin \theta \frac{u+u^\dagger}{2}\right)^N .
\label{ZnNcft}
 \eea
We stress that no approximation has been made in the above
procedure.

In the thermodynamic limit, $N\to\infty$ and $\theta \to 0$, one
has $\cos \theta \simeq 1$ and $\sin \theta\simeq \theta$. As a
result, the determinant in Eq.\ (\ref{ZnNcft}) may be
exponentiated:
 \be
Z_{n}(\theta)\equiv
\lim_{N\to\infty}Z_{n,N}\left(\frac{\theta}{N}\right)= \int_{{\rm
U}(n)} du\, \e^{i\frac{\theta}{2}{\,{\rm tr}\,} (u+u^\dagger)}\, .
\label{Zntdl}
 \ee

One can repeat
the above procedure for the BDI and CII classes.
The parametrization (\ref{param})
of the real and quaternionic Grassmannian
manifolds,
${\rm SO}(2N)/({\rm SO}(N)\times{\rm SO}(N))$ and
${\rm Sp}(4N)/({\rm Sp}(2N)\times{\rm Sp}(2N))$,
involves $N\times N$
real and quaternion--real matrices $Z$, instead of
complex.
The color--flavor transformation
trades the integrations over these `colored' variables $Z$
with the ones over `flavored' variables $u$ that are
antisymmetric and symmetric unitary matrices,
respectively (straightforward as they are,
such types of color--flavor
transformation have yet to be exhibited explicitly
in the literature,
to the best of our knowledge).
Accordingly, the integration domain of the
transformed partition function (\ref{ZnNcft}) becomes
antisymmetric unitary matrices (${\rm U}(2n)/{\rm Sp}(2n)$)
and symmetric unitary matrices (${\rm U}(2n)/{\rm O}(2n)$),
with the rest being unaltered.
Inclusion of non-zero $\nu$ is straightforward
by considering a rectangular $Z$. It merely shifts the
the power of the determinant in the measure (\ref{meas})
by $\nu$, and modifies
Eq.\ (\ref{Zntdl}) into
Eq.\ (\ref{aftercft})
in the thermodynamic limit.

For the B--D and C or DIII and CI classes, one utilizes
the color--flavor transformation between
the orthogonal or symplectic group and
the associated dual symmetric space parametrized by
antisymmetric or symmetric complex matrices, respectively
\cite{Zir1,Zir98,NN}.

\section{Generalized Color--Flavor Transformation}
\label{app2}

Kaneko \cite{Kan} (see also Yan \cite{Yan}) has derived the
following remarkable integral identity:
  \bea
Z_{n,N}(t)&=& \frac{1}{S_N(\Lambda_1+n,\Lambda_2,\l)}
\int\limits_0^1\prod_{i=1}^N
\left(dy_i\,y_i^{\Lambda_1}(1-y_i)^{\Lambda_2} (y_i-t)^n
\right)
\left( \Delta_N(y)^2 \right)^{\l}
\nonumber\\
&=& \frac{1}{S_n(V_1,V_2,1/\l)} \int\limits_{\mathcal{C}}
\prod_{a=1}^n \left(dx_a\,x_a^{V_1}(1-x_a)^{V_2} (1-t x_a)^N
\right)\left( \Delta_n(x)^2 \right)^{1/\l} \, ,
                                                   \label{gcft}
  \eea
where the constants are related as
  \be
V_1 = \frac{\Lambda_1+\Lambda_2+2}{\lambda}+N-2\, ; \hskip 1cm V_2
=  -\frac{\Lambda_2+n}{\lambda}-N\,  .
  \ee
The normalization constant $S_k(a,b,c)$ is given by the Selberg
integral
\begin{equation}
      \label{zN}
      S_k (a,b,c) =
      \prod_{j=0}^{k-1}\frac{\Gamma(a+1+c j)
        \Gamma(b+1+c j)
        \Gamma(1+c(j+1))}
      {\Gamma(a+b+2+c (k+j-1))\Gamma(1+c)}\, .
\end{equation}
The integration contour $\cal{C}$ encircles the cut between
$x_a=0$ and $x_a=1$.
In our case (cf. Eq.\ (\ref{Zn}))
  \be
\Lambda_{1,2}  = \l_{1,2} - \frac{1}{2}\, , \hskip 1cm
V_1 = \frac{\l_1+\l_2+1}{\l}+N-2\, ; \hskip 1cm V_2 =
   -\frac{\l_2+n-1/2}{\l}-N\, .
  \ee
We call the identity~(\ref{gcft}) fermionic replica ``generalized
color--flavor transformation''. Indeed,  for the RMEs values of
the parameters ($\lambda=1/2,1,2$ and special values of
$\lambda_{1,2}$)
Eq.\ (\ref{gcft}) essentially coincides with the fermionic replica
version of Zirnbauer's color--flavor transformation
\cite{Zir1,Zir2}, after proper parametrization of the symmetric
space elements and integration out of the irrelevant angles.
The
questions whether there is a geometrical ``dual pair'' interpretation of
Eq.\ (\ref{gcft}) for arbitrary parameters, and whether there is a
supersymmetric (say for rational $\lambda$) or bosonic replica
analogue, are currently open.

In the large--$N$ limit we collect all the terms having the
$N$--th power into $\exp[-N\sum_{a=1}^n S(x_a,t)]$, where
\begin{equation}
      \label{stat}
       S (x,t) = - \log(1-t x)- \log x + \log(1-x)\,  .
\end{equation}
We then look for the stationary points of the ``action'' $S(x)$
given by solutions of $\partial_x S =0$. A simple algebra gives
for the stationary points
\begin{equation}
      \label{statsol}
      x_\pm = 1\pm i \sqrt{\frac{1-t}{t}}  \, .
\end{equation}
We then magnify the vicinity of $t=0$ by introducing $\theta$ as $
t=\sin^2 (\theta/(2N)) \simeq \theta^2/(2N)^2$  and changing the
integration variable $x_a$ to $\phi_a$ as
  \be
x_a=1-i\sqrt{\frac{1-t}{t}}\,\e^{i \phi_a}\, .
  \ee
The two saddle points (\ref{statsol}) are at $\phi_a=0, \pi$.
Taking the limit $N\to\infty$, one obtains for the action $N
S(x_a,t) \to -i\theta \cos\phi_a$. A straightforward algebra
yields
  \be
\left( \Delta_n(x)^2 \right)^{1/{\lambda}}\,  \rightarrow \,
\theta^{-\frac{n(n-1)}{\lambda}} \prod_{a=1}^n
\e^{i\frac{n-1}{\lambda}\phi_a} \left[ \prod_{a>b}^n
\sin^2 \left(\frac{\phi_a-\phi_b}{2}\right)\right]^{{1}/{\lambda}}\, .
  \ee
The closed contour $\mathcal{C}$ in the $x_a$--plane can be taken to
be a circle, so that $\phi_a\in [0,2\pi]$. As a result, one
obtains Eq.\ (\ref{nfold1}) of the main text.

\end{document}